\documentclass[a4paper,12pt]{article}
\usepackage[pctex32]{graphics}

\begin{document}

\topmargin 0pt \oddsidemargin 0mm
\newcommand{\beq}{\begin{equation}}
\newcommand{\eeq}{\end{equation}}
\newcommand{\beqa}{\begin{eqnarray}}
\newcommand{\eeqa}{\end{eqnarray}}
\newcommand{\fr}{\frac}

\begin{titlepage}
\begin{flushright}
INJE-TP-06-01,~gr-qc/0603051
\end{flushright}

\vspace{5mm}
\begin{center}
{\Large \bf Phase transition between the BTZ black hole and AdS
space} \vspace{12mm}

{\large   Yun Soo Myung \footnote{e-mail
 address: ysmyung@inje.ac.kr}}
 \\
\vspace{10mm} {\em Institute of Mathematical Science, Inje
University,
Gimhae 621-749, Korea\\
 Institute of Theoretical Science,
University of Oregon, Eugene, OR 97403-5203, USA}
\end{center}
\vspace{5mm} \centerline{{\bf{Abstract}}}
 \vspace{5mm}
In three dimensions,  a  phase transition  occurs between the
non-rotating BTZ black hole and the massless BTZ black hole.
Further, introducing the mass of a conical singularity,  we show
that a transition
 between the non-rotating BTZ black hole
and thermal AdS space is also possible.
\end{titlepage}
\newpage
\renewcommand{\thefootnote}{\arabic{footnote}}
\setcounter{footnote}{0} \setcounter{page}{2}

\section{Introduction}
Hawking's semiclassical analysis for the black hole  radiation
suggests that most of information in initial states is shield
behind the event horizon and is never back to the asymptotic
region far from the evaporating black hole\cite{HAW1}. This means
that the unitarity is violated by an evaporating black hole.
However, this conclusion has been debated ever
since\cite{THOO,SUS}. It is closely related  to the information
loss paradox which states the question of whether the formation
and subsequent evaporation
 of a black hole is unitary. One of the most urgent problems in the black
hole physics is to resolve the unitarity issue.

Maldacena proposed that the unitarity can be restored if one takes
into account the topological diversity of gravitational instantons
with the same AdS  boundary in three-dimensional
gravity\cite{MAL}.  Actually, three-dimensional gravity\cite{BTZ}
is not directly related to the information loss problem because
there is no physically propagating degrees of freedom\cite{CAL}.
If this gravity is part of string theory\cite{HH}, the AdS/CFT
correspondence\cite{MGW} means that the black hole formation and
evaporating process should be unitary because its boundary can be
described by a unitary CFT. Recently, Hawking has withdrawn his
argument on information loss and suggested  that the unitarity can
be preserved  by extending Maldacena's proposal to
four-dimensional gravity system\cite{HAW2}.

 We remark an interesting
phenomenon in the AdS black hole thermodynamics. There exists the
Hawking-Page transition between AdS-Schwarzschild black hole and
thermal AdS space in four dimensions\cite{HP}.   Some authors have
proposed  that this transition is also possible  in
three-dimensional spacetimes: transition  between the non-rotating
BTZ black hole and  thermal AdS space\cite{KPR,KPR1}. Recently the
author has shown that there is no the first-order Hawking-Page
transition between the non-rotating BTZ black hole and thermal AdS
space\cite{myung}, by comparing it with the phase transition
between AdS-Schwarzschild black hole and thermal AdS space.

In this letter, we show that a  phase transition occurs
 between the non-rotating BTZ black hole and the massless
BTZ black hole.  If one introduces the mass of a conical
singularity,  a transition
 between the non-rotating BTZ black hole
and thermal AdS space is also possible. We use the off-shell
$\beta$-function which measures the mass of a conical singularity
at the event horizon, and the off-shell free energy which is used
to study the growth of the off-shell black hole.

We start with the non-rotating ($J=0$) BTZ black hole described by
the line element \beq \label{1eq2} ds^2_{NBTZ}=
-\Big[\fr{r^2}{l^2}-\mu\Big]dt^2
+\fr{dr^2}{\fr{r^2}{l^2}-\mu}+r^2d\theta^2
 \eeq
which possesses a continuous mass spectrum from $M=
\fr{\mu}{8G_3}$ to the massless AdS black holes ($M=0$) with
different topology: \beq \label{2eq2}
ds^2_{MADS}=-\frac{r^2}{l^2}dt^2+\frac{l^2}{r^2}dr^2+r^2d
\theta^2,\eeq where we find a degenerate event horizon at the
origin of the coordinate ($r=0$). Also the AdS spacetime is
allowed  by the line element\beq \label{3eq2} ds^2_{TADS}=
-\Big[1+\fr{r^2}{l^2}\Big]dt^2
+\fr{dr^2}{1+\fr{r^2}{l^2}}+r^2d\theta^2.
 \eeq
In this work  we consider three interesting
cases\cite{CLZ,myung1}. i) The non-rotating BTZ black hole (NBTZ)
is given by $M=r_+^2/8G_3l^2$ and $T_{H}=r_+/2\pi l^2$ with the
horizon radius $r_+=l\sqrt{\mu}$. ii) The massless BTZ black hole
(MBTZ) with $M=T_{H}=0$ is called the spacetime picture of the RR
vacuum state. iii) The thermal AdS spacetime (TADS)  is determined
by $M=-1/8G_3$ and $T_H=0$. This case corresponds to the spacetime
picture of the NS-NS vacuum state\cite{MS}. Although the
thermodynamic properties of TADS and MBTZ are nearly the same,
 their Euclidean topologies are quite different: TADS (MBTZ) are
topologically non-trivial (trivial). The TADS has a
non-contractible $S^1$ at $r=0$, while the MBTZ is contractible
 but it has a conical singularity at the event horizon ($r=0$).

In $d\ge 4$ case, the Hawking-Page phase transition occurs between
the Schwarzschild-AdS black hole and thermal AdS space. In this
case, there exists a minimum temperature at $r_+=r_0$. We have two
solutions: for $r_+<r_0$, the unstable black hole  with the
negative heat capacity; for $r_+>r_0$, the stable black hole with
the positive heat capacity.  Even though the unstable solution is
thermally unstable, it is important as the mediator of phase
transition from thermal AdS to AdS black hole.

\section{Transition between MBTZ and NBTZ}
However, for Chern-Simons black holes (NBTZ case), the situation
is quite different from the case of the Schwarzschild-AdS black
hole\cite{CTZ,BCM}. The NBTZ could be thermally equilibrium with
the heat reservoir at any temperature $T$. To show this, we
introduce the on-shell free energy (energy) and heat capacity
(entropy) as \beq \label{4eq2}
F^{on}_{NBTZ}=-E_{NBTZ}=-\frac{r_+^2}{8G_3l^2},~~
C_{NBTZ}=S_{NBTZ}=\frac{\pi r_+}{2G_3}. \eeq A condition for the
thermal equilibrium is given by $T=T_{H}$. Then we always have a
stable NBTZ at $r_s= 2 \pi l^2 T$ without the minimum temperature.
A positive heat capacity ($C_{NBTZ}>0$) means that the NBTZ is a
thermally stable system, irrespective of any size of the black
hole. This point contrasts to the case of the Schwarzschild-AdS
black hole. It is obvious that the NBTZ with $T_H=0$ leads to the
MBTZ case  \beq \label{5eq2}
F_{MBTZ}=E_{MBTZ}=C_{MBTZ}=S_{MBTZ}=0. \eeq On the other hand,
thermodynamic quantities for thermal AdS space are given by \beq
\label{6eq2}
F_{TADS}=E_{TADS}=-\frac{1}{8G_3},~~C_{TADS}=S_{TADS}=0. \eeq

\begin{figure}[t!]
   \centering
   \includegraphics{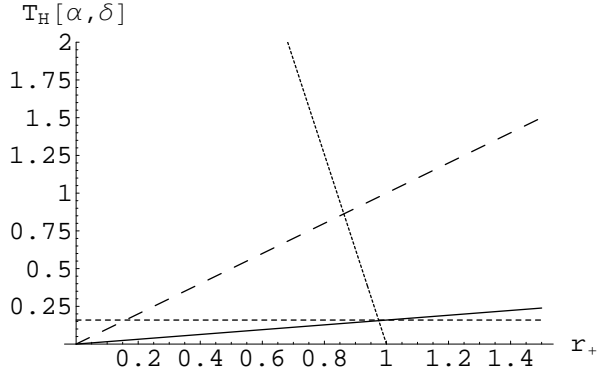}
\caption{ The temperature picture of a cool (off-shell) black hole
growth  in a hotter heat bath at $T=T_c=1/2\pi$ (small dashed
line).
  Solid line shows a plot of the  increasing temperature $T_{H}$ of a cool
   black hole with $l=1$. Large  dashed line  indicates the off-shell parameter
   $\alpha(r_+,T_c)$. Dotted  line denotes the deficit angle $\delta(r_+,T_c)$.
   In this case we have
a saddle point (stable NBTZ) at
$r_s=1(\alpha=1,\delta=0,T_H=T_c)$. } \label{fig1}
\end{figure}

\begin{figure}[t!]
   \centering
   \includegraphics{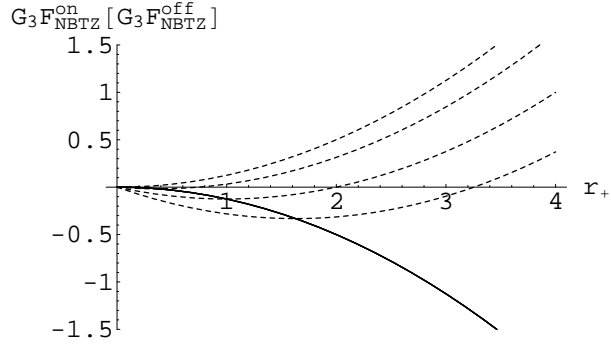}
\caption{The on-shell versus off-shell free energy.
   The solid line represents the on-free
energy $F^{on}_{NBTZ}(r_+)$ in the units of $G_3$ and $l=1$, while
the dashed line denotes the off-shell free energy
$F^{off}_{NBTZ}(r_+,T)$ for four different temperatures: from the
top down, $T=0, 0.059, T_c(=0.159), 0.259$. At each saddle point
$r_+=r_s$, we have $F^{out}_{NBTZ}=F^{on}_{NBTZ}$.} \label{fig2}
\end{figure}

In order to study the phase transition clearly, we have to
introduce the generalized (off-shell) free energy \beq\label{7eq2}
F^{off}_{NBTZ}=E_{NBTZ}-T\cdot S_{NBTZ}. \eeq Also the off-shell
parameter $\alpha$ and the deficit angle $\delta$ take the forms
\beq \label{8eq2} \alpha(r_+,T)=\frac{T_H}{T},~~\delta(r_+,T)=2
\pi(1-\alpha). \eeq As is shown in Fig. 1, $\alpha$ is zero at
$r_+=0$ and it is one at $r_+=r_s$ with $l=1$. On the other hand,
$\delta$ has the maximum of $2\pi$ at $r_+=0$ and it is zero at
$r_+=r_s$. This means that the near horizon geometry at $r_+=0$ is
the narrowest cone with the shape ($\prec$), while its geometry at
$r_+=r_s$ is a contractible manifold ($\subset$). In this sense
$r_+=0$ is not a saddle point.  We have $0 <\delta<2\pi$ between
$r_+=0$ and $r_+=r_s$ and thus we have a cone singularity at the
event horizon ($<$). Using $\alpha$, we can rewrite the off-shell
free energy as \beq\label{9eq2}
F^{off}_{NBTZ}(r_+,T)=-F^{on}_{NBTZ}\Big[1-\frac{2}{\alpha}\Big]
\eeq with the corresponding Euclidean action
 \footnote{At appendix D
of Ref.\cite{KPR1}, there is a slightly different approach to this
free energy. The on-shell action is given by $I^{on}_{NBTZ}=
F^{on}_{NBTZ}/T_H=-\frac{\pi r_+}{4G_3}$ and a contribution from
the conical singularity is $I_{cs}=-\frac{r_+\delta}{4G_3}$. The
total gravitational action is then:
$I_g=I^{on}_{NBTZ}+I_{cs}=\frac{\pi r_+}{4G_3}\alpha-\frac{\pi
r_+}{2G_3}=-I^{on}_{NBTZ}\alpha\Big[1-\frac{2}{\alpha}\Big]=
F^{off}_{NBTZ}/T$. Here we prove $I_g=I^{off}_{NBTZ}$. If this
conical deficit is created by a Euclidean point particle of mass
$M_{pp}$, we include its action $I_{pp}(=-I_{cs})=-\pi^2 l^2 T
\alpha(\alpha-1)/G_3$ as a counter term. Then the total action
leads to the on-shell action : $I_{tot}=I^{on}_{NBTZ}+I_{cs}
+I_{pp}=I^{on}_{NBTZ}$.} $I^{off}_{NBTZ}=F^{off}_{NBTZ}/T$.
 At $\alpha=1(r_+=r_s)$, we recovers $F^{off}_{NBTZ}=F^{on}_{NBTZ}$.
We confirm this from the operation  \beq\label{10eq2}
\frac{\partial F^{off}_{NBTZ}}{\partial r_+}=0 \to T=T_{H} \to
F^{off}_{NBTZ}=F^{on}_{NBTZ}. \eeq In this sense the off-shell
(off-equilibrium) free energy becomes the on-shell free energy at
the  saddle point of $r_+=r_s=2\pi l^2T>0$. Further, we obtain the
$\beta$-function from the definition \beq\label{11eq2}
\beta_{NBTZ}(r_+,T) \propto \frac{\partial
I^{off}_{NBTZ}}{\partial r_+}=-\frac{{\cal C}_{NBTZ}}{6
l}\delta(r_+,T), \eeq where the ${\cal C}_{NBTZ}$-function is
related to the central charge on the boundary CFT.  In this case,
it is just a constant ${\cal C}_{NBTZ}=3l/2G_3=c$. Further
Eq.(\ref{11eq2}) means that the $\beta$-function measures the
deficit angle $\delta$ mainly.

 At
this stage, we introduce an assumed picture of the phase
transition in three dimensions. A  phase transition may occur at
$T=T_c=1/2\pi l(r_+=l)$ between NBTZ and MBTZ\cite{myung}. As is
shown in Fig. 2, at $T=0$, the MBTZ is a saddle point as the
ground state.  For $T>0$,  we have $F^{off}_{NBTZ}(r_+)<0$ at the
saddle point $r_+=r_s$ so that a stable NBTZ is more probable than
the MBTZ.  Thus it is possible to flow from the  MBTZ to the NBTZ
along the path provided by the off-shell black hole
configurations. At $T=T_c$, the situation is the same. This case
is depicted in Fig. 3. The off-shell black holes can be modeled by
the metric Eq.(\ref{1eq2}) with fixed $T$ and varying $0< r_+ <
r_s$, and a conical singularity at the event horizon. This differs
from the Hawking-Page transition where the unstable black hole
plays an important role of the mediator from thermal AdS to AdS
black hole. Here is no such a mediator.   Hence there is no the
Hawking-Page like transition in three dimensions. This states the
censorship for the Hawking-Page transition in three dimensions.
Since, in the canonical approach, the free energy corresponds to
the effective potential, the transition between the MBTZ with
 and NBTZ  may be regarded as
the tunneling process.

\begin{figure}[t!]
   \centering
   \includegraphics{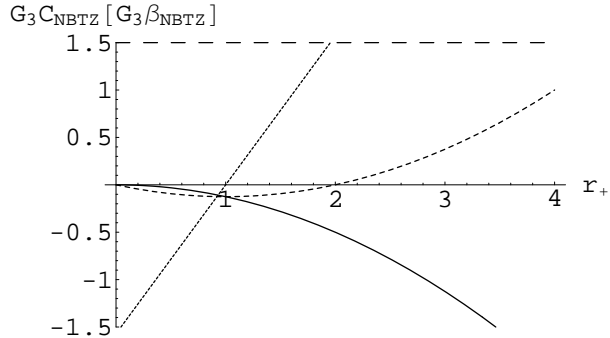}
\caption{The large dashed line denotes the ${\cal
C}_{NBTZ}=3l/2G_3=c$ as the central charge on the  CFT boundary.
The dotted line represents the off-shell $\beta$-function
$\beta_{NBTZ}(r_+,T_c)$, which measures the mass of a conical
singularity. The solid line denotes the on-shell free energy
$F^{on}_{NBTZ}(r_+)$, while the small dashed line shows the
off-shell free energy $F^{off}_{NBTZ}(r_+,T_c)$. At the junction
point of $r_+=r_s=1$, one has
$F^{on}_{NBTZ}=F^{off}_{NBTZ},~\beta_{NBTZ}=0$. This point is a
stable NBTZ which comes out from the off-shell approach.}
\label{fig3}
\end{figure}

On the AdS side, we check whether or not the Cardy-Velinde formula
is satisfied with this picture. To obtain this formula of
$S_{NBTZ}=\frac{2\pi l}{d-2}\sqrt{E_c(E_{NBTZ}-E_c)}$\cite{Ver},
we have to define the Casimir energy $E_c=2E_{NBTZ}-T_HS_{NBTZ}$.
However, it turns out  \beq \label{CE} E_c=0.\eeq Also considering
the boundary topology of $S^1$ leads to $E_c=0$ because it is
locally flat. Thus we no longer use this formula to show a
relation between entropy and energy in three dimensions.

On the CFT side, we introduce the well-known Cardy formula in two
dimensions \beq \label{CA}
S_{CFT}=2\pi\sqrt{\frac{c}{6}L_0}+2\pi\sqrt{\frac{\bar{c}}{6}\bar{L}_0}
\eeq with $c=\bar{c}=3l/2G_3$ and $L_0=\bar{L}_0=M_{NBTZ}l/2$.
Here the radius of $S^1$ is set to be $\rho=1$. Then we establish
the AdS/CFT correspondence for the entropy:
$S_{CFT}=S_{NBTZ}$\cite{myung1}. Finally we note that  this model
does not satisfies a higher-dimensional relation of $E_c \propto
c$ because of $E_c=0$.

\section{Transition between TADS and NBTZ}

In three dimensions, one has a mass gap between
 MBTZ  with $M_{MBTZ}=0$ and TADS with $M_{TADS}=-1/8G_3$.
  A conical singularity
interpreted as a point mass source would  be introduced to explain
this. For this purpose, we use  the relation of
$I_{cs}=-\frac{r_+\delta}{4G_3}\equiv 4 \pi r_+ M_{cs}$. Then the
mass of a conical singularity at the event horizon is given
by\cite{KPR1} \beq M_{cs}(r_+,T)=-\frac{1}{8 G_3}\frac{\delta}{2
\pi}=-\frac{1}{8 G_3}\Big(1-\alpha\Big).\eeq
 This is
closely parallel to the point particle at the event horizon:
$I_{pp}=\frac{r_+\delta}{4G_3}\equiv 4 \pi r_+ M_{pp}$ with
$M_{pp}=\frac{\delta}{16 \pi G_3} =-M_{cs}$. Here we obtain
another relation $M_{cs}=\beta_{NBTZ}/4\pi$ between mass and
$\beta$-function.
  The branch of $-1/8G_3 \le M_{cs}<0$
 is allowed only to a collection of off-shell black holes with a conical singularity
for $0\le r_+<r_s$. In this section  we use the mass (energy)
$M_{cs}$ instead of the mass of black hole itself.

Furthermore, we introduce a new  energy  and free energy which are
based on the Horowitz-Myers conjecture for the AdS
soliton\cite{HM}. This implies  that the soliton with a negative
energy can be taken as the thermal background.  We note that for a
three-dimensional AdS space, the flat AdS black hole and spherical
AdS black hole are the same because their horizons are one
dimension. Thus, the three-dimensional AdS solition is just the
thermal AdS space\cite{SSW}. Then we can calculate the new energy
and free energy with respect to the soliton background (TADS)
using the standard regularization scheme:
 \beq\tilde{E}(r_+)=\frac{1}{8G_3}\Big[1+\frac{r_+^2}{l^2}\Big],~
\tilde{F}^{on}(r_+)=F^{on}_{NBTZ}-F_{TADS}=\frac{1}{8G_3}\Big[1-\frac{r_+^2}{l^2}\Big].
\eeq This leads to \beq
\tilde{F}^{off}(r_+,T)=F^{off}_{NBTZ}(r_+,T)-F_{TADS}.\eeq  The
new energy of $\tilde{E}=E_{NBTZ}-E_{TADS}$ is always positive
with respect to the TADS. We have
$\tilde{F}^{on}=\tilde{F}^{off}=1/8G_3 $ but $M_{TADS}$ is found
to be $ M_{cs}(0,T_c)=-1/8G_3$ at $r_+=0$. On the other hand, at
the saddle point $r_+=r_s$, we have
$\tilde{F}^{on}(r_+)=\tilde{F}^{off}(r_+,T_c)=M_{cs}(r_+,T_c)=0$.
This is depicted in Fig. 4. At $T=T_c$, the transition from the
TADS to the NBTZ is possible. For $T<T_c$, the TADS dominates,
while for $T>T_c$, the NBTZ dominates because of
$\tilde{F}^{off}(r_+=r_s)<0$. There is a change of dominance at
the critical temperature $T=T_c$.

Therefore, if one considers the mass of a conical singularity, we
can  connect the TADS with the NBTZ  using the off-shell free
energy approach. In this way, we could accommodate the TADS with a
negative mass and free energy within our picture.

\begin{figure}[t!]
   \centering
   \includegraphics{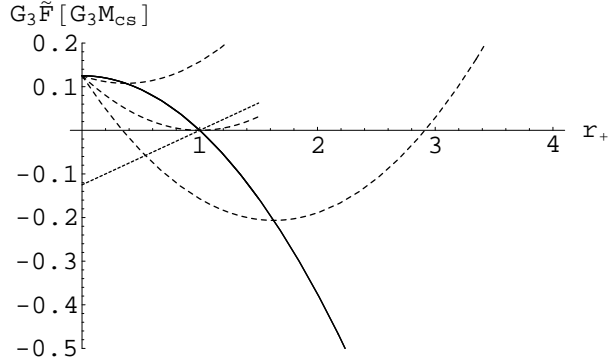}
\caption{  The solid line denotes the on-shell free energy
$\tilde{F}^{on}(r_+)$, while the  dashed lines show the off-shell
free energy $\tilde{F}^{off}(r_+,T)$ for three different
temperatures: from the top down, $T=0.059(<T_c),~T_c=0.159,~
0.259(>T_c)$. These are shifted from $F^{on}_{NBTZ}$ and
$F^{off}_{NBTZ}$ by $+1/8$. The dotted line represents the mass of
a conical singularity $M_{cs}(r_+,T_c)$.} \label{fig4}
\end{figure}

Alternatively, if one includes quantum fluctuations, there exits a
possibility that the MBTZ is not the end of the Hawking
evaporation and the end might be the TADS\cite{LO}.

 Consequently, the
transition between the MBTZ and NBTZ is possible to occur. This
does not belong to the first-order Hawking-Page transition because
it is not a genuine process of a black hole nucleation mediated by
an unstable black hole. Furthermore, if one introduces the mass of
a conical singularity and the Horowitz-Myers conjecture,   a
transition
 between the TADS
and NBTZ  is also possible.

\section*{Acknowledgement}
 This work was
supported by the Korea Research Foundation Grant
(KRF-2005-013-C00018).

\end{document}